\newcommand{\paperdate}{\today}
\newcommand{\papertitle}{Multimodal urban mobility and multilayer transport networks}
\newcommand{\paperkeywords}{human mobility, multilayer networks, transport networks, complex systems, urban data science, science of cities}

\RequirePackage[l2tabu,orthodox]{nag}   % warn if using any obsolete or outdated commands
\documentclass[11pt,onecolumn]{article} % document style

\usepackage[T1]{fontenc}                % output T1 font encoding (8-bit) so accented characters are a single glyph
\usepackage[utf8]{inputenc}             % allow input of utf-8 encoded characters
\usepackage{crimson}                    % document's serif font, in the style of minion pro
\usepackage{helvet}                     % document's sans serif font, helvetica
\usepackage{amssymb}
\usepackage{bm}

\usepackage[strict,autostyle]{csquotes} % smart and nestable quote marks
\usepackage[USenglish]{babel}           % automatically regionalize hyphens, quote marks, etc
\usepackage{microtype}                  % improves text appearance with kerning, etc

\usepackage{abstract}                   % allow full-page title/abstract in twocolumn mode
\usepackage{authblk}                    % footnote-style author/affiliation info
\usepackage{booktabs}                   % better looking tables
\usepackage{caption}                    % custom figure/table caption styles
\usepackage[final]{draftwatermark}      % watermark paper as a draft
\usepackage{geometry}                   % configure page dimensions and margins
\usepackage{graphicx}                   % better inclusion of graphics
\usepackage{hyperref}                   % hypertext in document
\usepackage{rotating}                   % rotate wide tables or figures on a page to make them landscape
\usepackage{setspace}                   % configure spacing between lines
\usepackage{titlesec}                   % custom section and subsection heading
\usepackage{url}                        % make nice line-breakble urls
\usepackage{natbib}    % author-year citations w/ bibtex, including textual and parenthetical  [numbers]
\setcitestyle{authoryear,open={(},close={)}}

\usepackage{adjustbox}
\usepackage{multirow}
\usepackage{subfigure}
\usepackage{eurosym}

\usepackage{amsmath}

\usepackage{lipsum} %Include loremipsum text

\usepackage{color}

\makeatletter
\def\BState{\State\hskip-\ALG@thistlm}
\makeatother


\graphicspath{{./figures/}}

\titleformat{\section}{\normalfont\sffamily\large\bfseries\color{black}}{\thesection.}{0.3em}{}
\titleformat{\subsection}{\normalfont\sffamily\small\bfseries\color{black}}{\thesubsection.}{0.3em}{}
\titleformat{\subsubsection}{\normalfont\sffamily\small\bfseries\color{black}}{\thesubsubsection.}{0.3em}{}

\titleformat{\paragraph}[runin]{\normalfont\bfseries}{}{}{}[]

\hypersetup{
	pdfauthor={Natera Orozco, L., Alessandretti, L., Saberi, M., Szell, M., Battiston, F.},
	pdftitle={\papertitle},
	pdfsubject={\papertitle},
	pdfkeywords={\paperkeywords},
	pdffitwindow=true,         % window fit to page when opened
	breaklinks=true,           % break links that overflow horizontally
	colorlinks=false,          % remove link color
	pdfborder={0 0 0}          % remove link border
}

\begin{document}

\title{\papertitle}
\date{\paperdate}
\author{Luis Guillermo Natera Orozco$^1$, Laura Alessandretti$^2$,\\ Meead Saberi$^3$, Michael Szell$^{4,5}$, Federico Battiston$^1$\footnote{Corresponding author: BattistonF@ceu.edu}}

\affil[]{\small
		 $1$) Department of Network and Data Science, Central European University, 1100 Vienna, Austria\\
		 $2$) Technical University of Denmark, DK-2800 Kgs. Lyngby, Denmark\\
		 $3$) School of Civil and Environmental Engineering, UNSW Sydney, NSW 2032, Sydney, Australia\\
		 $4$) NEtwoRks, Data, and Society (NERDS), IT University of Copenhagen, 2300 Copenhagen, Denmark\\
		 $5$) Complexity Science Hub Vienna, 1080 Vienna, Austria\\}

\maketitle
\begin{onecolabstract}
\noindent
Transportation networks, from bicycle paths to buses and railways, are the backbone of urban mobility. In large metropolitan areas, the integration of different transport modes has become crucial to guarantee the fast and sustainable flow of people. Using a network science approach, multimodal transport systems can be described as multilayer networks, where the networks associated to different transport modes are not considered in isolation, but as a set of interconnected layers. Despite the importance of multimodality in modern cities, a unified view of the topic is currently missing. Here, we provide a comprehensive overview of the emerging research areas of multilayer transport networks and multimodal urban mobility, focusing on contributions from the interdisciplinary fields of complex systems, urban data science, and science of cities. First, we present an introduction to the mathematical framework of multilayer networks. We apply it to survey models of multimodal infrastructures, as well as measures used for quantifying multimodality, and related empirical findings. We review modelling approaches and observational evidence in multimodal mobility and public transport system dynamics, focusing on integrated real-world mobility patterns, where individuals navigate urban systems using different transport modes. We then provide a survey of freely available datasets on multimodal infrastructure and mobility, and a list of open source tools for their analyses. Finally, we conclude with an outlook on open research questions and promising directions for future research.


\vspace{0.5cm}
\end{onecolabstract}

\section*{Introduction}

Urban mobility takes place across a range of transport modes. The most basic mode is walking, allowing individuals to cover short distances. With the historic growth of cities and transport technologies, new forms of mobility have emerged that are able to cover larger distances, keeping the growing city a coherent unit: from horsecar, bicycle, streetcar, to rail, bus, and automobile. These different modes of transport fulfil different roles that cater to the heterogeneity of urban trips which unfold across a range of different spatial scales, from individual buildings to city blocks, neighbourhoods, up to large urban agglomerations \citep{batty2006hierarchy, alessandretti2020scales}. Due to their heavy-tailed distribution \citep{brockmann2006scaling, gonzalez2008understanding, song2010modelling, alessandretti2017multi}, the majority of urban trips are short and can therefore be taken by foot~\citep{varga2016further}. The second largest fraction of trips is medium distance optimally taken by bicycle or bus. Finally, the smallest fraction of urban trips is long distance, for which travel via metro/rail or via car is reasonable~\citep{varga2016further}. It is therefore important for any large enough city to have a balanced mix of modes that reflects the distribution of how far its citizens need to move. 

The subject of this review, multimodal mobility (or combined mobility), is then the idea to combine multiple modes in one trip. On one hand this is happening naturally just by the existence of multiple modes - for example, a trip with a bus includes walking to and from a bus station. On the other hand, multimodal mobility can also be intentionally designed, such as bicycletrains \citep{geurs2016multi} or shared mobility solutions to the last mile problem \citep{shaheen2016mobility}. Given the different ranges, densities, and frequencies of transport modes, their combination can thus be advantageous. \textbf{This is the potential of multimodal mobility: Combining multiple transport modes promises to offer the benefits of all modes while avoiding their weaknesses.} A multimodal transportation system offers this benefit by connecting different transportation modes through interfaces (e.g. stations) that facilitate transfers between the distinct services \citep{vannes2002design}.

\subsection*{Potential and challenges of multimodal mobility}

The biggest potential of multimodal mobility is to improve urban sustainability. In the 20th century, the now conventional car-centric approach to transport planning has emerged aiming to minimize travel time for individual motorized transport \citep{banister2005unsustainable}. However, with the further growth of cities, it became clear that a monolithic focus on cars is unsustainable. The economic burden of automobility alone is \euro\,500 billion yearly in the EU while walking and cycling provides a yearly benefit to society worth \euro\,90 billion due to positive health effects \citep{gossling2019social}. Sustainable urban transport development therefore views mobility as a means to provide \emph{people} social access to the city rather than an optimized traffic flow for vehicles. In this modern view, all transport modes co-exist in a hierarchy that prioritizes walking, cycling, and transit over other forms of transport \citep{banister2005unsustainable,gossling2016urban}. This co-existence naturally provides opportunities for combinations. However, designing a well-functioning multimodal transport system comes with challenges.

\begin{figure*}[t!]
	\centering
	\includegraphics[width=\textwidth]{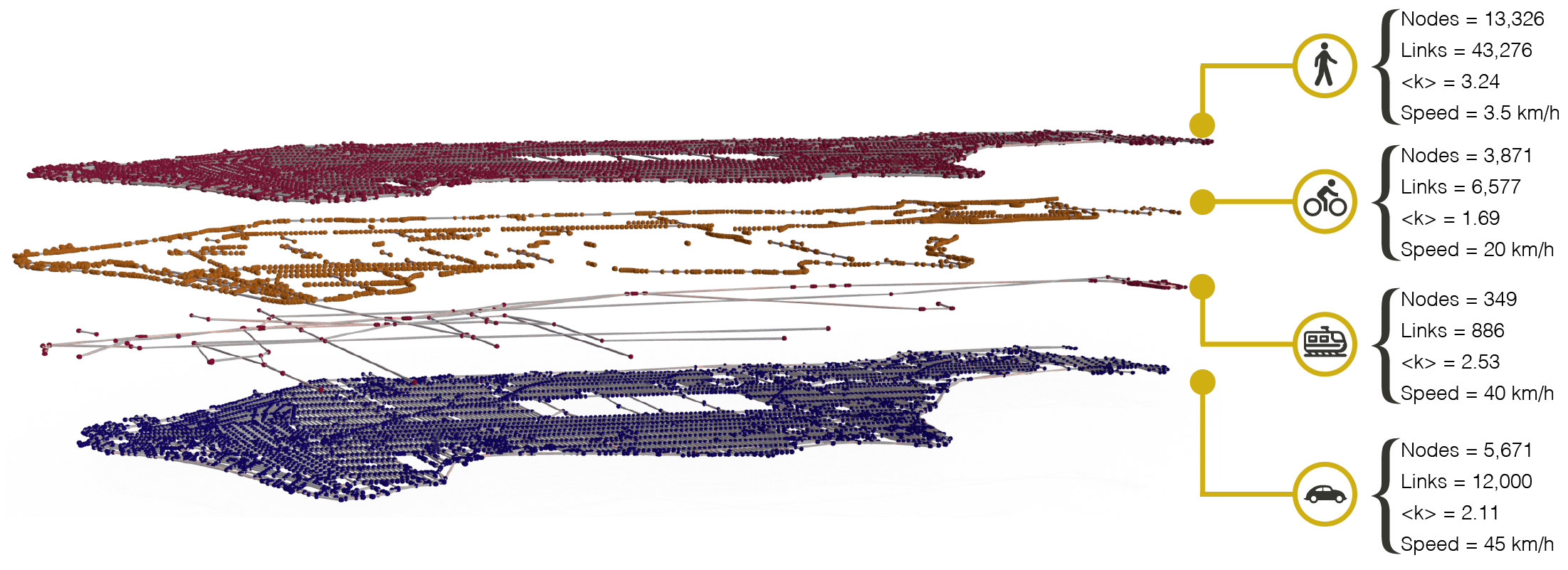}
	\caption{\textbf{Multiplex network representation of Manhattan, New York City,} with four layers of transport infrastructure (pedestrian paths, bicycle paths, rail lines, and streets), with data from OpenStreetMap~\citep{OpenStreetMap}. \textbf{(Right)} Network information for each layer, number of nodes, links and average degree $\langle k \rangle$. Figure adapted from \cite{natera2020multimodal}.}
	\label{fig:Manhattan}
\end{figure*}

Apart from deep political challenges related to car dependence \citep{mattioli2020political}, each mode of transport has its own underlying infrastructure network and schedule of operation that has unique structural and dynamical properties depending on its function. An example of this structural complexity is provided in Figure~\ref{fig:Manhattan} showing the interconnected transport system of Manhattan given by the interplay of a pedestrian, a bicycle, an underground and a car network, characterised by different size, connectivity and velocity. Considering private transport, the network of pedestrian sidewalks and walking areas must be at least as dense as the bicycle network, as almost every location in the city should be reachable by walking but not necessarily by bicycle. Similarly, considering public transport, a bus network will generally be more dense and more frequently operated than an urban rail network which serves larger distances.

The new multimodal nature of large cities raises further planning challenges. From the traveller's view, choosing convenient travel solutions within multimodal systems requires to process large amounts of information \citep{gallotti2016limits}. From the transport provider's perspective, multimodal systems are more difficult to manage, due to the large number of aspects that have to be taken into account in order to achieve synchronization between different agencies and modes \citep{barthelemy2016structure}. In practice, these difficulties have created novel business opportunities for both proprietary and public stakeholders, e.g. mobile phone applications for multimodal urban mobility \citep{willing2017intermodal}, and a plethora of research questions which are the subject of this review.

\subsection*{Aim and structure of this Review}

The primary aim of this article is to offer a comprehensive review of multimodal transportation and mobility research focusing on recent approaches from complexity science. Within this framework a city is understood as a complex system~\citep{batty2013new,lobo2020urban}, and its transport infrastructures, such as streets, sidewalks, bicycle lanes and public transport, are modelled as networks~\citep{barthelemy2011spatial}. The field of Network Science has extensively studied such spatial networks, first from a single-layer and more recently with a multilayer perspective, with special attention to  transportation networks~\citep{lin2013complex,barthelemy2011spatial,ding2019application}, focusing on topological properties~\citep{jiang2004topological,cardillo2006structural,barthelemy2008patterns,batty2008size,barthelemy2011spatial,strano2013comparative,louf2014typology,boeing2020multiscale}, centrality metrics~\citep{crucitti2008centrality,boeing2018planarity,kirkley2018structural}, and growth processes~\citep{makse1995growth,strano2012elementary,szell2021gub}. Other topics include the impact of the street networks on pedestrian volume \citep{hajrasouliha2015connectivity}, accessibility and vitality of cities~\citep{denadai2016death,biazzo2019accesibility,natera2019walkability}, resilience of transportation networks~\citep{baggag2018resilience,ferretti2019resilience,natera2020growth}. These recent approaches can be seen as the beginning of emerging fields like a Science of Cities or Urban Data Science which exploit new large-scale urban datasets with quantitative tools from physics, geoinformatics, and data/network science \citep{batty2013new,resch2019hds}.

In this review, we survey the literature on urban multimodal mobility, and on urban transportation infrastructure as multilayer networks. Here, we focus on the primal approach to networks \citep{Porta2006primal} where streets and mobility infrastructure constitute the network links, and intersections (bus stops, subway stations, etc.) constitute the nodes of the network. Other approaches such as space syntax are not common in complexity science.

The remainder of this review is arranged as follows. Section~\ref{sec:multilayernetworks} introduces the mathematical concept of multilayer networks. In Section~\ref{sec:multimodalinfrastructures} we discuss research on urban transport infrastructures, including their empirical characterization (see Subsection \ref{subsec:charachterising_infrastructure}) and theoretical modelling (see Subsection \ref{subsec:modeling_infrastructure}) as multilayer networks. In Section~\ref{sec:multimodalmobility} we instead focus on mobility and navigation across these multimodal systems. In Section
~\ref{sec:datatools} we cover the relevant open datasets and the main software tools which can be used to analyse multimodal transportation systems. We conclude with an outlook and a summary of open questions for the research community in Section~\ref{sec:conclusions}.

\section{Multilayer networks: a framework for multimodality}\label{sec:multilayernetworks}

Over the last decades, networks have emerged as a versatile tool to understand, map and visualise the interconnected architecture of a wide range of complex systems~\citep{albert2002statistical, dorogovtsev2002evolution, newman2003structure, boccaletti2006complex}, in particular spatially embedded ones~\citep{barthelemy2011spatial, barthelemy2018morphogenesis}. Formally, a network, or graph, $\mathcal G = (\mathcal N, \mathcal L)$ consists of a set of nodes $\mathcal N$, and a second set $\mathcal L$ of edges, describing connections among unordered pairs of elements of the first set. This information can be stored into an adjacency matrix $\bm A = \{a_{ij}\}$, where $i=1, \dots, N$ are the nodes, and $a_{ij}=1$ if there is a link between nodes $i$ and $j$, $a_{ij}=0$. In transportation systems~\citep{lin2013complex}, nodes can represent the stations of a network, and links direct connections between them. Two approaches for constructing network links have been proposed \citep{barthelemy2011spatial}: in the first one, known as L-space representation, two nodes are connected if they are consecutive stops in a given transportation route; in the second one, known as P-space representation, two nodes are connected
if there is at least one transportation route between them. Interestingly, we note that while this is the most used notation, such wording is the result of a mistake, as L and P should in principle be exchanged, as they refer respectively to logical and physical network. 

The adjacency matrix can also include weights $\bm W = \{w_{ij}\}$, where $w_{ij}$ are positive real numbers, for instance describing how strongly connected two nodes are. For spatial systems, weights are often taken as the reciprocal of the distance between two nodes, or the time it takes to travel from one to another, i.e. $w_{ij}=1/d_{ij}$ or $w_{ij}=1/t_{ij}$.

More recently, network scientists have put effort into characterising the structure of systems which are formed by different interconnected networks. Such interconnected structures are natural for transportation systems but also for social and biological networks. Think for instance of the largest transportation hubs in worldwide cities, where stations are routinely served by bus, underground and railway infrastructures.
Indeed, most urban transportation systems systemically rely on the interplay between different means of transportation. These systems can be conveniently described by \textit{multiplex} or \textit{multilayer} networks. Here we introduce the \textit{vectorial} formalism for multilayer networks~\citep{boccaletti2014structure, battiston2014structural}, widely used in most papers on multimodal transportation. An alternative description can be provided by a more mathematically involved \textit{tensorial} framework~\citep{dedomenico2013mathematical, kivela2014multilayer}.  

In multilayer networks, links of different types, describing for instance a different mean of transportation, are embedded into different \textit{layers}. Each layer $\alpha$, $\alpha = 1, \ldots, M$, is described by an adjacency matrix 
$\bm W^{[\alpha]} = \{w_{ij}^{[\alpha]}\}$. In a multimodal urban transportation network with three mobility infrastructure layers, $\alpha=1$ can represent the bus network, $\alpha=2$ the underground network, and $\alpha=3$ the urban railway network, for example. The full transportation system $\mathcal M$ can be described as $\mathcal W = \{\bm W^{[1]}, \ldots, \bm W^{[M]}\}$. Nodes $i=1, \dots, N$ are labeled in the same order in all networks. 

In the case of transportation networks, simply identifying nodes of different networks as the same station might not provide the most complete description of the multimodal network. Take for instance the largest stations in mega-cities, like King's Cross - St. Pancras in London, Grand Central Station in New York or Hongqiao transportation hub in Shanghai. All of these are identified by a unique location (node index) $i$ across the different transportation layers. Yet, switching from one mean of transportation to another within the same station might require a non-negligible fraction of time and effort, given the complexity and size of the overall infrastructure. 

For this reason, it is often relevant to complement the description of the \textit{intra-layer} connections present in the system, with \textit{inter-layer} links associated to the cost, spatial distance or time required to switch layers. Inter-layer links between layers $\alpha$ and $\beta$ at a node $i$ can be encoded through the inter-layer matrix $C_i=\{c_i^{[\alpha \beta]} \}$, and all such inter-layer connections can be stored in the vector $\bm C = {C_1, \ldots, C_N}$. In this case, the full multiplex structure of the system is described by taking into account both intra-layer and inter-layer connectivity, hence $\mathcal M = (\bm W, \bm C)$. Inter-layer links may be neglected for many measures focusing on diversity~\citep{battiston2014structural}, as well as correlations~\citep{nicosia2015measuring} across the layers of the systems, relevant to assess the different roles and geographical spanning of the different mean of transportations of a multilayer network. 

Multilayer networks are a natural framework for multimodal transport networks. Indeed, the concept of multilayer networks in the transportation engineering field goes back to the 1970’s when \cite{dafermos1972traffic} proposed a formulation for the traffic assignment problem for multiclass-user networks. Since then, the term ``multilayer'' has been interchangeably used with ``multiclass'', ``multimodal'', and ``multiuser''. Shortly after, \cite{sheffi1978transportation} proposed the notion of ``hypernetwork'' that was later redefined as ``supernetwork'' \citep{sheffi1985urban} in which decision-making can be modelled as a route selection over a multilayer or multimodal network. In the field of network science, one of the pioneering works introducing the framework and concept of ``layered complex networks''~\citep{kurant2006layered} explicitly focuses on the case of transportation systems, where a first layer encodes the physical infrastructure of the system, and the second one describes the flows on such infrastructure. Other early works on the topic also dealt with interconnected systems at the worldwide level, focusing on different modes of transport such as the multiplex airline networks~\citep{cardillo2013emergence}.
Noticeably, multimodal infrastructures seem to possess exclusive characteristics different from other multilayer networks. For instance, when tools to assess the redundancy of the different layers are considered, transportation networks are often found to be irreducible~\citep{dedomenico2015structural}. Differently from many biological systems, where layers often duplicate information to guarantee the interconnected system a high level of robustness, the layers of a multiplex transportation systems are purposedly engineered to be different, in order to maximise efficiency~\citep{latora2001efficient}. As a byproduct of this feature, multimodal systems are also often highly fragile~\citep{buldyrev2010catastrophic} and sensitive to disruptions or failures of a single infrastructure~\citep{dedomenico2014navigability}. For the reader interested in further material on the topic, we refer to the early reviews~\citep{boccaletti2014structure, kivela2014multilayer} and textbook~\citep{bianconi2018multilayer} covering the field. \cite{aleta2019multilayer} provide a more recent eye-bird view of the field. A thorough review of the measures and models used to analyse such systems can be found in \cite{battiston2017new}, whereas \cite{dedomenico2016physics} give a theoretical overview of spreading and diffusive processes on such systems. In the following sections of this review, we focus on findings of more direct relevance to the research community working with multimodal transportation and urban mobility. The division of related research across these two themes is not meant to be rigid, but rather serves as a indication of the core topic treated in the different works.

\section{Multimodal infrastructures}\label{sec:multimodalinfrastructures}

As cities grow and add different transportation modes, understanding the transportation infrastructure and its interconnected nature is crucial to capture patterns of urban mobility. Since the 1950s, fields ranging from Architecture to Urbanism and Transport Planning have grown a large body of literature studying the structure of cities and their transport systems. With the growth of the Complex Systems and Network Science fields, new models have been developed to study the complexity behind urban systems, and specifically mobility infrastructure.

Multimodal urban infrastructures can be represented as multilayer networks, in which each layer $\alpha$ represents a mobility infrastructure (e.g. subway, light railway, bus service, pedestrian or bicycle infrastructure), the set of nodes $\mathcal{N}$ are locations (e.g. bus stops, intersections, subway stations), and the set of edges $\mathcal{L}$ in layer $\alpha$ are the infrastructure links between nodes in the same layer (e.g. subway lines, bus routes, bicycle lanes, see also Section \ref{sec:multilayernetworks}). Modeling infrastructures is of great importance for understanding how urban systems work, and for the design of new sustainable mobility options. 

In the following we describe recent findings related to how transportation layers are coupled and grow. First, we review the main complex systems models of urban transportation. Then, we describe empirical findings related to the transport infrastructures.

\subsection{Modeling multimodal infrastructures}\label{subsec:modeling_infrastructure}
 
The design of efficient single-layer transportation networks is a classical problem in the domain of transport optimization. As such, it is widely studied and often formulated as a bilevel mathematical problem, where in the upper-level the traffic planner makes decisions regarding management of the system, and in the lower level users choose route, travel mode, origin and destination of their travel in response to the upper-level decision. The design of multimodal transportation network, instead, has been less studied \citep{farahani2013review}. In fact, multimodality brings new challenges to the transportation network design problem, including issues related to the integration of the street network with public transit and active transportation networks (walking and cycling networks) in which travellers can choose to take multiple modes for a single journey \citep{zhang2014design,huang2018multimodal}.

One of the first contributions to the modeling of multimodal urban infrastructures was provided by \cite{decea2005equilibrium} who pointed out that most of the models from the transport community \citep{boyce1994introducing,boyce2002sequential,decea2007transit} used to plan and simulate the effects of new transportation options failed to consider congestion associated with transit modes. This shortcoming is particularly relevant when the models are used for infrastructure development and predict transportation equilibria in future years. 

In the work by \cite{decea2005equilibrium}, the authors take the road network as the base layer. There, links have an average operating cost that takes into account different mobility options (e.g. cars, taxi, etc.). For every public transportation mode a new layer is defined, with their unique nodes and links. For these layers, the cost function of the public transportation links depends on the combined effect of travel, waiting, and transfer time. The model considers the existence of combined trips (e.g. car/metro, bus/metro, etc) and looks for an equilibrium condition under the assumption that every user chooses her route to minimize their average operation cost (Wardrop’s first principle). This means that at equilibrium, only non-congested routes have a minimum cost, while those without flow represent a more costly option. Travel time might be affected by the interplay of different transportation means. For instance, vehicle flow over the road network may induce longer travel times in the bus network. Similarly, a passenger might decide not to take the subway if it is too crowded. \cite{decea2005equilibrium} show that their model is able to find the equilibrium in the trips between origin and destination in a toy model, and can be successfully applied to real-world scenarios. For instance, a rich version of this model (which comprised of 13 user classes, 11 transport modes, and 450 zones) successfully informed the planning of the new metro line 5 in Santiago (Chile).

A similar philosophy was deployed by \cite{li2007parkride} who considered a transportation infrastructure of cars, combined walk-metro paths, and park-and-ride. Differently from the previous work in which the model takes in consideration the availability of routes, here parking availability and time spent by car commuters while looking for parking, which can be considerable \citep{shoup2017high}, is considered explicitly. The focus here is on the interplay and impact of park-and-ride (P\&R) schemes to encourage users to switch from car travel to subway and public transportation options when traveling to the cities' central area. The model proposed by \cite{li2007parkride} considers the effects of traffic conditions on travel demand, and incorporates elastic demand into the model to capture commuters’ responses to traffic congestion and availability of parking supply. The responses of a user include the decision to switch to another transportation option, or to not make the trip at all. For the public transportation layer, the model also takes in consideration discomfort that may result from crowded subways. Through numerical simulations, \cite{li2007parkride} found that it is possible to reach an equilibrium control which prevents the emergence of traffic jams in the city center by implementing a suitable P\&R scheme, which looks at the combined effect of cost at the P\&R sites, parking availability in the city center, as well as metro fares and frequency.

While in the above works the interplay of different transportation modes was introduced, they were not yet modeled as a multilayer network. One of the first applications of this new modeling framework was presented by \cite{morris2012transport} who developed a toy model to couple different transport modes, according to the following rules. First, \textit{N} nodes are placed at random within the unit circle, mimicking a spatial configuration typical of many cities, and are connected by a Delaunay triangulation. Second, to simulate another transport mode, a second layer is generated by drawing a subset of the previously generated nodes and a second Delaunay triangulation. Finally, the two layers are coupled with interlayer links when a node is present in both layers. Using this toy model and a simulated Origin-Destination matrix, \cite{morris2012transport} investigated how fragile the network is to changes in supply and demand. They found that increasing travel speed in one layer tends to concentrate trips in the fastest layer, and also produces congestion in the nodes where is possible to change transport mode.

Similar results have been reported for coupled random networks \citep{gao2017comprehensive}, as well as scale-free networks \citep{zhuo2011traffic} where a similar modelling approach is used to mimic a real-world transport scenario \citep{du2016physics}. In their work, \cite{du2016physics} used a two-layer traffic model, where one layer provides higher transport speed than the other, and applied a Particle Swarm Optimisation algorithm to optimize the transport system capacity and reduce congestion both in synthetic and real-world multimodal networks. A more detailed coverage of the findings can be found when discussing betweenness centrality and interdependence in Section~\ref{subsec:charachterising_infrastructure}. 

\cite{gil2014configuration} proposed to use open data from OpenStreetMap to model the multimodal infrastructure network of a given city as a combination of three layers. The first layer is the street network, where the nodes are intersections and links are streets. This layer, accounting for private transportation, was again the reference layer with respect to the other transportation modes in the system, i.e.~all other layers have to be connected to and interact with it. The second layer is the public transport layer that represents the stations as nodes. It links the stations whenever there is a public transport service between two stations. This layer is coupled to the street layer by the stations and their closest street intersection. \cite{gil2014configuration} also include in their model land use to measure urban accessibility. This framework was applied to the analysis of the Randstad city-region in the Netherlands. The model was tested under different parameters and layer combinations, measuring reachability of the different city areas through closeness and betweenness centralities. Comparison with ground truth data showed that betweeness centrality in the public transportation layer can be a good indicator of passenger flows.

\cite{aleta2017transportation} exploited one step further the richness of multilayer networks for transportation systems, highlighting two possible frameworks. In the first framework, each bus or metro line on its own could be considered an independent layer. This approach is useful to have a realistic model of human mobility which takes into consideration transfer times and synchronization between single trips. Yet, it does not allow to evaluate the importance of an entire transportation mode. This issue is resolved in the second framework. Here all the lines of the same mode are combined into a so-called \textit{superlayer}, which is fundamental to study the interdependence and resilience of the whole system. 

Using the previously described frameworks, \cite{aleta2017transportation} investigated the public transport systems of nine European cities. Following the first framework, the authors focused on some structural features of the emerging system such as the overlapping degree (sum of the node's degree in all layers \citep{battiston2014structural}). They found that public transport infrastructures have some universal properties, and that the maximum overlapping degree is quite similar in all the systems, even if the number of layers is different. This depends on the fact that networks are embedded in a physical space, hence imposing some bounds on the maximum number of links of each node structural constraints. Following the second framework, \cite{aleta2017transportation} investigated in detail the superlayers and found that -- suprisingly -- the nodes with the highest overlapping degree are not necessarily the ones with the highest superlayer activity \citep{nicosia2015measuring}. Indeed, some transportation modes (and in particular the bus layer) have a tendency for hubs which might be disconnected to the other transportation modes, leading to high overlapping degree but low multilayer activity. The prevalence of such hubs is relevant when considering the robustness of the whole system. As the study highlights, it is often easier to move a bus stop to a street nearby, even if it is a local hub where multiple lines stop, than solving a disruption in a subway station. \cite{aleta2017transportation} also assess the importance of the superlayers based on the number of shortest paths that make use of the superlayer, a measure that we characterize later in Section~\ref{subsec:charachterising_infrastructure}.

It is important to note that cities and their transportation systems are not static in time. This means new transport modes may be introduced or extended, such as when new bus/subway stops are added. Recently, to plan for open streets during the COVID-19 pandemic, \cite{rhoads2020planning} proposed an investigation of the relation between streets and sidewalks. First, the authors used percolation theory to examine whether the sidewalk infrastructure in cities can withstand the tight pandemic social distancing. They then proposed an algorithm that takes into consideration both the sidewalk and street layers while improving the sidewalk connectivity. Despite spatial constraints, \cite{rhoads2020planning} showed that it is possible to widen the sidewalks and improve the pedestrian connectivity with a minimum loss in the road network. In a similar fashion, \cite{szell2021gub} explored different growth strategies for retrofitting streets to bicycle networks. Exploring the topological limitations of various bicycle network growth processes, \cite{szell2021gub} found initially decreasing returns on investment until a critical threshold, posing fundamental consequences to sustainable urban planning: Cities must invest into bicycle networks with the right growth strategy, and persistently, to surpass a critical mass. Decreased directness for automobile traffic due to bicycle network growth was found which is a desirable effect from the perspective of urban sustainability and livability.

More in general, some works have focused on the growth and evolution of multilayer networks, such as \cite{nicosia2013growing}, \cite{kim2013coevolution}, and \cite{nicosia2014nonlinear}, generalising preferential attachment mechanisms in different ways \citep{barabasi1999emergence}. However, these works do not keep into account spatial constraints, and are not well suited to describe the evolution of spatial networks. For this reason, there is ample potential for future work to develop growth models for multimodal infrastructures, for instance considering densification and exploration, as previously done for street networks \citep{strano2012elementary}. An alternative view can be obtained by investigating the optimal growth and design of the multiplex structure of the different layers, as was done for the multilayer airline network composed by routes of different airline companies \citep{santoro2018pareto}. 

The models discussed in this section show how the multilayer network framework has been used to investigate the structure, function and vulnerabilities of a complex transportation system. In the next section we cover some measures to quantify the multiplexity of these structures and their interconnections.

\subsection{Characterizing multimodal infrastructures}\label{subsec:charachterising_infrastructure}

\begin{figure*}[t!]
 	\centering
 	\includegraphics[width=\textwidth]{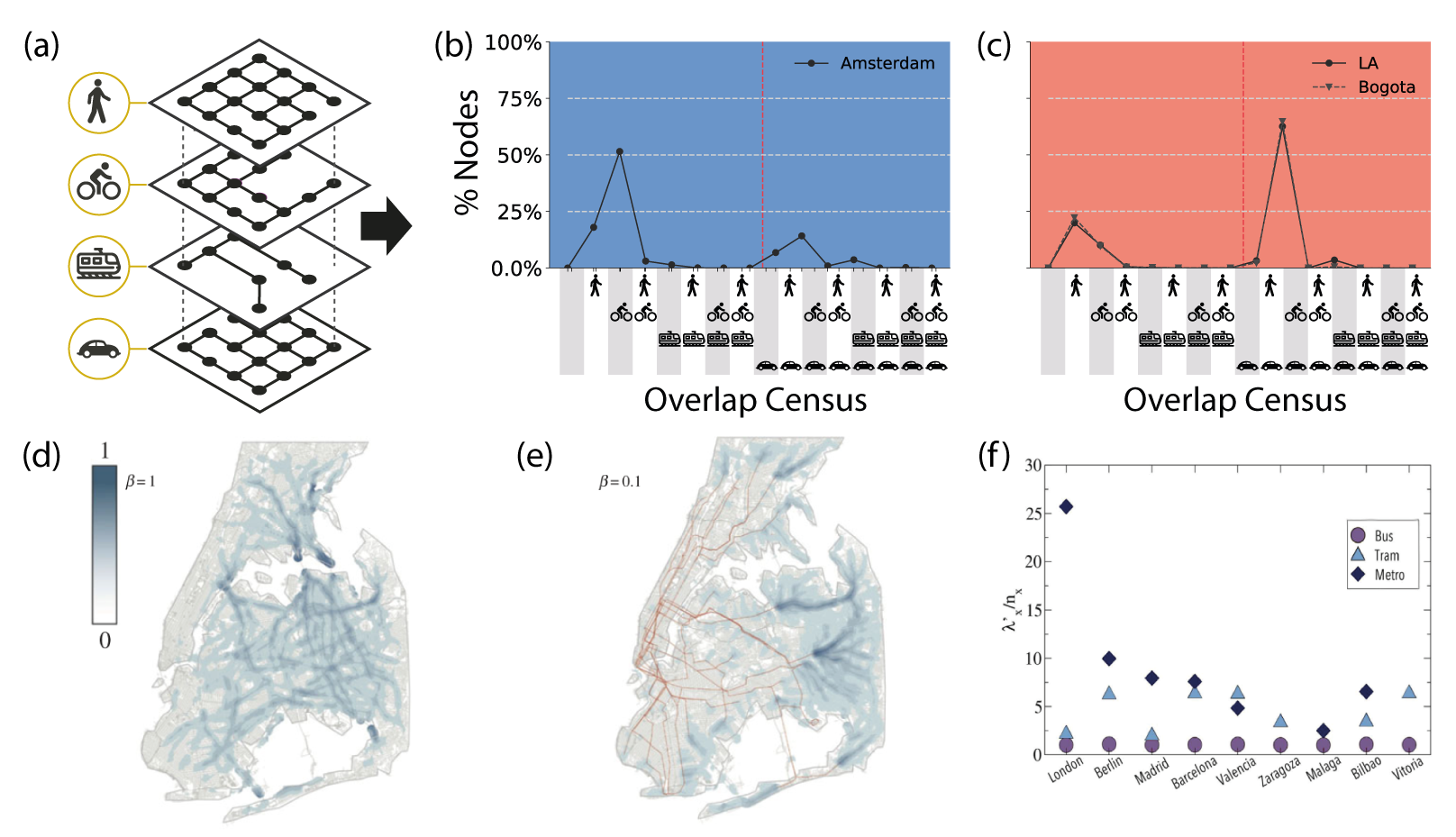}
 	\caption{\textbf{Structural measures of multimodality.} (a) Schematic of a city's multiplex transportation network. The information of the different layers is integrated in the overlap census (b). We show the case of Amsterdam (b, mostly active in non-motorized layers) versus Los Angeles and Bogota (c).
 	  Spatial distribution of betweenness centrality in New York. In (d) we show only the street layer. the subway layer is added in (e), showing a reallocation of centralities from avenues to metro stations. (f) Superlayer interdependency for multiple transportation networks. The metro layer comprises of most shortest paths in the system, making such transportation layer the most beneficial for high interdependence.
 	  Figures adapted from \cite{natera2020multimodal,strano2015features,aleta2017transportation}.}
 	\label{fig:Structure}
\end{figure*}

The empirical study of urban transportation infrastructure has revealed some of the structural properties of specific transport networks \citep{barthelemy2011spatial}. But how to quantify the effectiveness of their interconnections? In this section, we describe a number of measures that have been used to capture the multiplexity of multimodal infrastructures, such as the importance of different nodes, the system's resilience, and the similarity between layers.

\paragraph*{Overlap census.}\label{overlap}

Urban transportation networks present different degrees of multimodality and integration. How can we quantify such differences? The overlap census proposed by \cite{natera2020multimodal} is a method that helps to answer this question. Taking the multimodal mobility network of a city as an input, this measure calculates the fraction of nodes that are active in different multiplex configurations (Fig.~\ref{fig:Structure}~(a)). Given a multiplex transport network with $\mathcal{M}$ layers, the overlap census is formally defined as a vector with $(2^M)-1$ components, where each component stores the fraction of nodes that are reached by a given subset of the existing transport modes. For example, one component stores the fraction of nodes reachable by foot and bicycle but not by car, another the fraction of nodes that can be reached using any of the available transport modes (Fig.~\ref{fig:Structure}~(b-c)). Using the proposed metric, \cite{natera2020multimodal} investigated the different transportation profile of fifteen cities around the world, covering cities in multiple development stages, from Amsterdam and Copenhagen to Beihai or Detroit. By clustering together similar profiles, six clusters were identified which quantitatively capture different levels of development and sustainability.

The overlap census was not the first measure to compare similarities between multiplex networks and layers. Indeed, similar approaches were developed for multiplex networks more in general. For a detailed view see \cite{nicosia2015measuring} who proposed measures capturing nontrivial correlations in multiplex networks and models to reproduce those correlations. More recently, \cite{brodka2017similarity} also presented an overview of different metrics to compute similarities between layers in multiplex networks.

\paragraph*{Paths.} 
At a more global scale, multimodality is often associated with the ability of an agent to navigate the system by using the available transportation modes. Journey or route planning in a multimodal setting refers to finding the shortest route in a multilayer network when multiple transportation modes such as private car, public transit, walking, and cycling are combined in a single and integrated journey \citep{zografos2008algorithms,botea2013multi}. For this reason, the navigation of an agent in a transportation network can be measured through the available types of paths.

The literature on multimodal route planning has been mostly built around static multilayer networks with no time-dependent characteristics \citep{bast2016route}. Thus, a first possibility is to consider the quickest path, neglecting transfer and waiting times between transportation modes. This path is computed using the fastest speed associated to the edges, assuming a perfect synchronization between the different transportation modes. This is analogous to finding the shortest path between $i$ and $j$ in an aggregated weighted-single layer network, where the weight describing the time to travel between two nodes is the minimum among those offered by the different transportation options.

Yet, this approach often falls short in capturing real patterns of mobility. Indeed, transportation networks also have a temporal dimension that has to be taken into consideration. In order to find a path that allows for a change between two or more different transportation options we must find a time-respecting path \citep{gallotti2014efficiency}, defined as the shortest path between nodes $i$ and $j$ that considers the departures and arrivals constraints given by timetables. In order to find such time-respecting paths, \cite{huang2019route} proposed a time-expanded model to account for the dynamic nature of route planning in a multimodal network with fixed public transit and dynamic ridesharing vehicles. However, a comprehensive route planning algorithm that simultaneously considers multiple transportation modes in a truly time-dependent network is missing. Furthermore, for multimodal transportation networks, the walking transfer time between modes has to be taken into account when computing time-respecting paths. 

Besides time constraints, in order to find a viable path between $i$ and $j$, the logic of the proposed sequence of modes has to be assessed~\citep{battista1996path, lozano2001path}. For instance, in some cases, a path composed by subway-bus-subway-car-subway might solve the shortest path, but the presence of private transportation (car) as an intermediate option makes it an ``illogical path'' and thus an unlikely choice for a user. The validity of transportation sequences can in general be formalised in terms of cost associated to change of transportation mode.

Finding viable paths is one of the most important problems in urban transportation, as it has the potential to help users finding the most efficient paths in the city. \cite{lozano2001path} have proposed an efficient algorithm to find such paths when the agent establishes her limitations on the number of modal transfers.

The contribution of the different layers of a multiplex networks to shortest paths might be very unequal \citep{aleta2017transportation}. For instance, the rail systems (trams, subway) contribute to most of the shortest paths in a city, connecting distant points at a greater velocity and in straighter routes than bus or other transport modes. However, such layers have only few stations. Slower and more local transport modes often serve a complementary role, offering a deeper coverage of the city.

\paragraph*{Spatial outreach.}
The availability of different transportation modes, such as subways or tramways affects how easily it is to reach certain locations in the city. A way to measure this effect is to quantify the associated \textit{spatial outreach}~\citep{strano2015features}. The spatial outreach can be computed as the average distance from node $i$ to all other nodes in the same layer $\alpha$ that are reachable within a given travel cost $\tau$. Mathematically it is defined as follows:

\begin{equation}
    L_\tau(i)=\frac{1}{N(\tau)}\sum_{j|\tau_{m}(i,j)<\tau}d_e(i,j),
    \label{eq:outreach}
\end{equation}

where $d_e(i,j)$ is the distance between nodes $i$ and $j$, and $N(\tau)$ is the number of nodes reachable on the multilayer network within a travel cost $\tau$.

\cite{strano2015features} modified the average speed (traversal time of a link) in the layers of multiplex systems to measure their effects on the corresponding travel outreach. \cite{strano2015features} found that when the metro speed increases compared to the street layer speed, a clear area of high-outreach nodes emerges in the city center and around the nodes that have connections to the high-speed layer. In other words, as the velocity in layer $\beta$ increases, the nodes that are closer to the interchange nodes in layer $\alpha$ improve their accessibility, implying that a person can efficiently travel from this area to faraway places. This concept of travel outreach is similar to that of isochrones which quantify the accessible area from a given point within a certain time threshold, e.g.: What is the area that a user can reach traveling $x$ minutes, in any direction, from a given point? \cite{biazzo2019accesibility} used this approach to measure accessibility in different urban areas computing the isochrones as a combination of public transit and pedestrian infrastructure. With this method, scores were obtained that capture how well a city is served by the public transit and how accessible a specific area is to the rest of the city.

\paragraph*{Betweenness centrality and interdependence.}
The relevance of the nodes in a network is characterized by centrality scores. Additional to single layer networks, in multiplex networks the overall centrality of a location or station also depends on the interplay of the different transportation options.

The simplest way to define the centrality of a node is to measure its degree, or the number of locations directly connected to it. Such a measure, however, is not so relevant for systems embedded in space due to spatial constraints. For example, the number of possible connections of a given node (intersection) in the street layer is highly constrained by the physical space, as a single intersection can only have a limited number of intersecting streets/sidewalks. For this reason, other centralities than degree are typically used to assess the relevance of a location. One of such measures is the betweenness centrality~\citep{Freeman1977Centrality}, measuring the number of shortest paths passing via a given node. This measure is also called ``load'' and can be seen as the simplest proxy for traffic flow in the system as it assumes uniform demand between each pair of nodes. In the absence of explicit mobility data, betweenness centrality can be used as a proxy to assess the areas at risk to become overcrowded and to identify potential bottlenecks in the system. 

In multiplex networks, shortest paths go from one node to another able to pass through two or more layers. This effect can be quantified by measuring the interdependence of a given node $i$ as

\begin{equation}
    \lambda_i=\frac{1}{N-1}\sum_{j\neq i}\frac{\psi_{ij}}{\sigma_{ij}},
    \label{eq:coupling}
\end{equation}

where $\psi_{ij}$ is the number of shortest paths between $i$ and $j$ that use edges in two or more layers, and $\sigma_{ij}$ is the total number of shortest paths between $i$ and $j$~\citep{morris2012transport, battiston2014structural,strano2015features}. Node interdependence takes values in $[0, 1]$, with values close to 1 associated to a high coupling of the layers, while values close to $0$ mean that most of the paths from that node to other nodes go through just one layer. By taking the average over all nodes $\lambda = 1/N \sum_i\lambda_i$ it is possible to obtain a single score for the whole system. 

This interdependence measure can be modified to obtain a score for a specific layer~\citep{aleta2017transportation}. Then the layer interdependence for layer $\alpha$ is defined as:

\begin{equation}\label{eq:layer_interdependency}
    \lambda^{\alpha}=\frac{\sum_i\sum_{i\neq j}\psi_{ij}^{\alpha}}{\sum_i\sum_{i\neq j}\psi_{ij}},
\end{equation}

where $\psi_{ij}^\alpha$ describes the number of shortest paths between nodes $i$ and $j$ using two or more layers and such that at least one of them corresponds to layer $\alpha$.

When applying this measure to multimodal transport networks, \cite{aleta2017transportation} found that the metro and tram layers play an important role in concentrating shortest paths (Fig.~\ref{fig:Structure}~(f)). For Madrid, \cite{aleta2017transportation} found that more than 40\% of the trips have at least one link in the metro layer, even if the metro layer has only 241 nodes while the bus layer has 4590 nodes. 

As cities grow and new lines and transport modes are added into the mobility system, new interconnections between layers appear, changing the betweenness of the different nodes. \cite{ding2018traffic} studied how centralities evolved when the rail network of Kuala Lumpur grew from a tree-like structure to a more complex one. Their findings suggest that, as the network grows, the average shortest path in the multilayer network can decrease dramatically, especially as new nodes are able to serve as interchange between layers, thus enabling new shortest paths along the system.

The results from \cite{ding2018traffic} are in line with previous findings by \cite{strano2015features}, on how the subway layer affect the distribution of nodes centrality in London and New York. \cite{strano2015features} show that the introduction of new interconnected layers affects the congestion of the street layer. In fact, the presence of the subway layer allows to move traffic from internal routes and bridges to the terminal points of the subway system (Fig.~\ref{fig:Structure}~(e)) which might be used as interchange locations for suburban flows into the city center. Theoretical work by \cite{sole-ribalta2016congestion} confirms that one of the main drivers affecting traffic dynamics and congestion in multimodal transport networks is the interchange from the least to the most efficient layers.

\paragraph*{Resilience.}
Evaluating the robustness of a transport system under failures is an important task with practical implications in urban planning. Notably, multimodality affects significantly the resilience of transportation system \citep{dedomenico2014navigability}. 

In a single layer network, the disruption of an infrastructure, i.e.~the removal of a link, can make a station or a part of the city disconnected. For example, imagine a transit station in a single layer network: If the links connecting the station with the rest of the system are removed, the station is inaccessible. However, if such a station is part of a multimodal transportation network, it could still be accessed through other layers. To measure the impact of multimodality on resilience, \cite{dedomenico2014navigability} used random walks (see Sec.\ref{mobility_1}) to mimic trips among locations and investigated the coverage time in the London's transportation system under different scenarios, showing that the interconnected nature of the different transport modes dramatically enhances the overall system resilience to failure compared with the single layers. A similar approach was followed by \cite{baggag2018resilience}, where the coverage time of random walks was used to measure the robustness of the multimodal transportation networks of Paris, London, New York, and Chicago. To mimic realistic trips, Baggag et al. introduced several constrains on the complexity of the trips, for instance limiting the maximum number of transport mode changes. More recently \cite{ferretti2019resilience} used the multiplex framework to model Singapore's public transportation infrastructure and test its resilience against floods in the city in different scenarios, finding that the system is extremely resilient as it faces the first significant disruption only after the removal of $~50\%$ of it edges.

\section{Multimodal mobility}\label{sec:multimodalmobility}

Understanding urban travel is paramount for a range of real-world applications, including planning transportation~\citep{patriksson2015traffic} and designing urban spaces. Starting from the 1950s, a large body of literature in the fields of Geography and Transportation has studied how people move and use transportation technology. 

As the transport infrastructure becomes increasingly multimodal, modelling how individuals make travel decisions in complex interconnected networks is critical. In recent years, the scientific understanding of human mobility has dramatically improved, also due to the widespread diffusion of mobile-phone devices and other positioning technologies, which allowed to gather large-scale geo-localized datasets of human movements and develop increasingly realistic behavioural models. Concurrently, these recent developments had benefited by the dramatic growth of the fields of Complex Systems and Network Science, which brought together ideal tools to study interconnected systems. For a comprehensive review of the recent literature stream of Human Mobility see~\citep{barbosa2018human}. This new data-driven modeling framework for multimodal mobility was pioneered by \cite{kurant2006extraction}, who extracted data from public transportation timetables to characterize the mobility structure and traffic flow of a transportation network. Despite recent advancements, our understanding of multimodal mobility in urban systems remains limited, also due to the difficulties related to collecting comprehensive data across multiple transportation modalities. 

In this section we review the scientific literature on multimodal mobility. While our focus will be on multimodality, we will inevitably touch upon some of the concepts related more broadly to modelling of urban travel. In Section~\ref{mobility_1}, we briefly sumarize existing models, focusing on latest advances driven by the Complex Systems literature. In Section~\ref{mobility_2}, we review measures and empirical findings, with a focus on recent studies based on passively collected data sources. 

\subsection{Modeling urban mobility \label{mobility_1}}

Modeling travel in a multimodal system involves understanding how individuals make decisions in a constantly changing complex environment. The most common family of models for travel demand in the Geography and Transportation literature are the \emph{four-step models} proposing that each trip results from four decisions~\citep{mcnally2000four}: 1) whether to make a trip or not, 2) where to go, 3) which mode to use, 4) and which path to take. For simplicity, these steps have been largely considered as independent, sequential choices, and correspond to four modelling steps: trip generation, trip distribution, mode choice, and route assignment. See the review of \cite{mcnally2000four} for a comprehensive overview about this modeling approach.

In recent years, the field of Complex Systems has modeled travel behavior on multiplex networks using different approaches that we briefly review in this section. Complex Systems research has proposed novel individual \citep{song2010modelling, jiang2016timegeo, alessandretti2020scales} and collective \citep{simini2012universal,schlapfer2020hidden} models that capture well the first two aspects of travel behaviour: trip generation and trip distribution. In this review, we will focus largely on the last two of the four modelling steps, mode choice and route assignment, because they are the most relevant in the framework of multimodality. It is important to remark that, also due to the lack of empirical data on the mechanisms driving human navigation, many models rely on simplistic assumptions, for example that individuals are rational, homogeneous, or have unlimited knowledge. Research based on novel data sources will be key to develop mobility models on multilayer networks that include realistic elements such as limited knowledge and cognitive limitations.

\paragraph{Random walks.} The random walk is one of the most fundamental dynamic processes \citep{sole2016random} that has been widely studied in the Complex Systems literature as a prototypical model for numerous phenomena occurring upon networks, including human mobility. Importantly, in contrast to widely used models that assume individuals with global knowledge of the system thus choosing shortest routes \citep{wardrop1952road}, random walks assume that agents are only aware of the local connectivity at each node. A random walk on a graph is defined by a walker that, located on a given node $i$ at time $t$, hops to a random nearest neighbor node $j$ at time $t + 1$. In the case of multilayer networks, the walk between nodes and layers can be described with four transition rules accounting for all possibilities \citep{dedomenico2014navigability}: (i) $P_{ii}^{\alpha\alpha}$, the probability for staying in the same node $i$ and layer $\alpha$; (ii) $P_{ij}^{\alpha\alpha}$ the probability of moving from node $i$ to $j$ in the same layer $\alpha$; (iii) $P_{ii}^{\alpha\beta}$ the probability of staying in the same node $i$ while changing to layer $\beta$; (iv) $P_{ij}^{\alpha\beta}$ the probability of moving from node $i$ to $j$ and from layer $\alpha$ to $\beta$, in the same time step. These probabilities depend on the strength of the links between nodes and layers, e.g. the frequency of vehicles and the cost associated to switching layers. 

Despite their simple formulation, random walks provide fundamental insights to many types of diffusion processes on networks and allow to measure a network's dynamical functionality. For example, random walk processes were used to measure the navigability of multiplex networks \citep{dedomenico2014navigability}. 
To this end, one can measure the coverage of the multiplex network $\rho(t)$, defined as the average fraction of distinct nodes visited by a random walker in a time shorter than $t$ (assuming that walks started from any other node in the network), and describing the efficiency of a random walk in the network exploration:

\begin{equation}\label{coverage}
    \rho(t)=1-\\\frac{1}{N^2}\sum_{i,j=1}^{N}\delta_{i,j}(0)\text{exp}[-\mathbf{P}_j(0)\mathbb{P}\mathbf{E}_i^{\dagger}],
\end{equation}

where $\mathbf{P}_j(0)$ is the supravector of probabilities at time $t=0$, the matrix $\mathbb{P}$ accounts for the probability to reach each node through any path of length $1, 2, \dots, \text{or}\ t+1$, and $\mathbf{E}_i^{\dagger}$ is a supra-canonical vector allowing to compact the notation. \cite{dedomenico2014navigability} provided an alternative representation of equation \ref{coverage} building upon the eigendecomposition of the supra-Laplacian. \cite{dedomenico2014navigability} showed that the ability to explore a multilayer network is influenced by different factors, including the topological structure of each layer and the strength of interlayer connections and the exploration strategy. Further, they showed that the multilayer system is more resilient to random failures than its individual layers separately because interconnected networks introduce additional paths from apparently isolated parts of single layers, and thus enhance the resilience to random failures.

Random walks have further been used to assign a measure of importance to each node in each layer, by measuring the asymptotic probability of finding a random walker at a particular node-layer as time goes to infinity, the so-called \emph{occupation centrality} \citep{sole2016random}. \cite{sole2016random} provided analytical expressions for the occupation centrality in the case of multilayer networks.

\paragraph{Travel time minimization approaches.} 
At the other end of the spectrum, agents are assumed to have global (or nearly global) knowledge of the system.This is one of the most widely-used approaches in the transportation literature, rooted in Wardop’s user equilibrium principle~\citep{wardrop1952road} for traffic assignment. Under the user equilibrium principle, in a congested system, all agents choose the best route, e.g. no user may lower his transportation cost through unilateral action. \cite{uchida2005study,zhou2008dynamic,verbas2015dynamic} are among many of the studies that proposed formulations and solution algorithms for traffic assignment under user equilibrium in a multimodal system.

Complex Systems research has developed models where agents aim at minimizing their individual travel times in congested \citep{tan2014congestion,bassolas2020scaling,manfredi2018congestion,sole-ribalta2016congestion} or uncongested \citep{du2014traffic,du2016physics} networks. 

\cite{bassolas2020scaling} developed an agent-based models describing mobility of individuals through a multilayer transportation system with limited capacity. The routing protocol used by individuals for planning is adaptive with local information. In the absence of congestion, individuals follow the temporal optimal path of the static multilayer network calculated by the Dijkstra algorithm. If there are line changes, \cite{bassolas2020scaling} estimate besides the change walking penalty an additional waiting time of half the new line period (the real waiting time will be given by the vehicles location in the line when the individual arrives at the stop). An individual's route is only recalculated when a congested node, whose queue is larger than the vehicle’s capacity, is reached. The work investigates analytically (for simple networks) and via numeric simulations the robustness of the network to exceptional events which give rise to congestion, such as demonstration concerts or sport events. The study revealed that the delay suffered by travellers as a function of the number of individuals participating in a large-scale event obeys scaling relations. The exponents describing these relations can be directly connected to the number and line types crossing close to the event location. The study suggested a viable way to identify the weakest and strongest locations in cities for organizing massive events.

Similarly, \cite{manfredi2018congestion} introduced a limit to the nodes capacity of storing and processing the agents. This limitation triggers temporary faults in the system affecting the routing of agents that look for uncongested paths. 

Importantly, the assumption that individuals have global knowledge of the system and minimize travel time contrast recent findings in spatial cognition, showing that human spatial knowledge and navigation ability is limited \citep{gallotti2016limits,bongiorno2021vector}. For example, a recent study on pedestrian navigation made clear that path choices seem to be affected by the orientation of street segments along the route \citep{bongiorno2021vector}. Recent modelling approaches for single-layer networks \citep{manley2018exploring} incorporate these ideas in routing models where agents are characterized by bounded knowledge and limited rationality. Further research will be necessary to develop realistic multilayer routing models accounting for the limits of spatial cognition.

\subsection{Characterizing multimodal mobility \label{mobility_2}}

Traditionally, multimodal mobility models are calibrated using data from travel surveys: \emph{Revealed Preference} surveys retrieve actual travel information from the respondents, while \emph{Stated Preference} surveys expose the travelers to various hypothetical scenarios and record their choices~\citep{arentze2013travelers}. Studies based on survey data have provided insights into how multimodal travelers value aspects such as the different travel time components (in-vehicle time, walk time, access time, wait time...)~\citep{abrantes2011meta}, service quality~\citep{wardman2001review}, travel costs ~\citep{arentze2013travelers}, and heterogeneities across socio-demographic groups ~\citep{nobis2007multimodality}. Due to the high costs associated with data collection and inherent biases in self-reported data, these studies suffer of serious limitations, including small sample sizes, data inaccuracy and incompleteness \citep{chen2016promises,zannat2019emerging}. Covering empirical results from travel surveys is outside the scope of this review, and we refer the reader to \cite{arentze2013travelers} for a comprehensive introduction to the topic.

In recent years, the empirical research on Human Mobility has taken new directions. A growing body of literature has focused on quantitative descriptions of human movements from large, automatically collected data sources, such as mobile phone records, travel cards and GPS traces \citep{barbosa2018human}. In this section, we give an overview of recent empirical findings on multimodal mobility in the field of Complex Systems which focused on two important aspects: 1) the dynamics of public transport systems, whose study was driven by the availability of public transport data such as schedules and positions of stop and stations, and 2) individual multimodal behaviour driven by the availability of data collected using `smart travel cards' and GPS data. 

The existing empirical research on multimodal travel based on passively collected data-sources is far from being comprehensive. Most studies have focused on public transit, such that the interplay between public and forms of private transportation such as walking, driving and cycling has been poorly characterized. Further, several studies are based on public transport schedules instead of real-time data, thus neglect important effects deriving from congestion. The increasing availability of high-resolution GPS trajectories collected by individual mobile phones and sensors installed on private and public transport vehicles~\citep{barbosa2018human} will be key to fill these gaps in the literature.

\subsubsection{Public transport systems dynamics} 
Over the last decade, the availability of detailed public transport schedules shared by public transport companies (see also Section~\ref{sec:datatools}) has allowed to better estimate travel times and characterize transport systems. 

\paragraph{Efficiency} To satisfy the demand of large number of individuals while reducing energy and costs, multilayer transport systems must achieve high efficiency. One aspect concerns the \emph{synchronization} between the network layers, because the more layers are synchronized, the less users have to wait for vehicles. The synchronization inefficiency $\delta(i,j)$~\citep{gallotti2014efficiency,barthelemy2016structure} for nodes $i$ and $j$ can be measured as the ratio of the time-respecting travel time $\tau_t(i,j)$, which accounts for walking and waiting times and the fact that the speed of vehicles varies during the day, and the minimal travel time $\tau_m(i,j)$, assuming that vehicles travel at their maximum speed and that transfers are instantaneous:

\begin{equation}
    \delta(i,j)=\frac{\tau_t(i,j)}{\tau_m(i,j)}-1
\end{equation}

\begin{figure}[t]
     \centering
     \includegraphics[width=\textwidth]{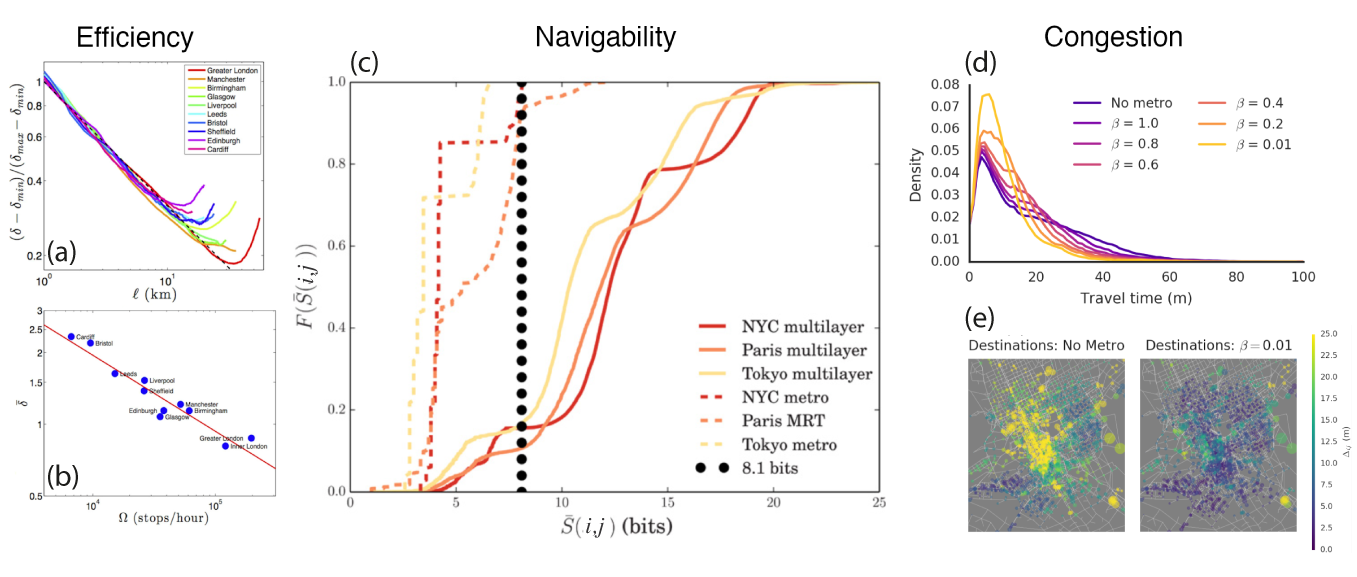}
    \caption{\textbf{Characterizing Public Transport Systems Dynamics.} \textbf{Efficiency} (a) The synchronization efficiency $d(i,j)$ between two nodes $i$ and $j$ against the path length $\ell(i,j)$ for different public transport systems in the UK. (b) The average synchronization efficiency $\overline \delta$ against the total number of stop-events per hour $\Omega$ (e.g. the number of times a vehicle stops) for different public transport systems in the UK (blue circles). The red line corresponds to  $\delta \sim \Omega^{-\mu}$, with $\mu=0.3$. \textbf{Navigability.} (c) The cumulative distributions of the information $S(i,j)$ needed to travel between source $i$ and target $j$ for the multilayer public transport networks of NYC, Paris, and Tokyo (solid lines). Most of the trips require more information than the cognitive limit $S_{max} \approx 8.1$ (dashed line) found for single layer systems. \textbf{Congestion.} The construction of a planned metro system in the city of Riyadh impacts congestion: (d) The probability distribution of travel times for different values of the ratio $\beta$ between the travel speeds of the street layer (car) and the planned metro layer. (e) Heatmap displaying the congestion impact $\Delta_{id}$ (where $i$ are origins and $j$ are destinations) aggregated over destinations $j$. The aggregates $\Delta_{i}$ can be interpreted as the expected impact of removing one driver who works at a given location from the streets. Results are shown in the absence of a metro system (left) and when the speed ratio $\beta=0.01$ (right). Figures adapted from \cite{gallotti2014efficiency,gallotti2016limits,chodrow2016congestion}.}
    \label{Panel2}
 \end{figure}

Using the synchronization inefficiency, \cite{gallotti2014efficiency} showed that, on average in the UK, $23\%$ of travel time is lost in connections for trips with more than one mode. Interestingly, across several urban transport system in the UK, the synchronization efficiency $\delta(i,j)$ obeys the same scaling relation with the path length $\ell(i,j)$:
\begin{equation} \label{eq:deltaSynchronization}
    \delta(i,j) \approx \delta_{\textit{min}}+\frac{\delta_{\textit{max}}-\delta_{\textit{min}}}{\ell(i,j)^v},
\end{equation}
with $v \approx 0.5$ and where $\delta_{\textit{max}}$ and $\delta_{\textit{min}}$ are the maximum and minimum values of $\delta(i,j)$ for a given urban transport system (Fig.~\ref{Panel2}-a).

Further, \cite{gallotti2014efficiency} have shown that the average synchronization inefficiency $\overline{\delta}$ for a given urban system follows: 
\begin{equation}
   \delta \sim \Omega^{-\mu}
\end{equation}
where $\Omega$ is the total number of stop-events per hour (e.g. the number of times a vehicle stops), and $\mu = 0.3 \pm 0.1$ (Fig.~\ref{Panel2}-b).

Other studies focused on the efficiency in terms of ability to satisfy users demand. \cite{alessandretti2016user} introduced a method based on non-negative matrix factorization to compare the network of commuting flows and the public transport network. This methodology, applied to various public transport systems in France, showed that, while in Paris the transportation system meets the overall demands, it does not in smaller cities where people prefer to use a car despite having access to fast public transportation. \cite{sui2019publictransport} proposed three topological metrics to quantify the interaction between public transport network and passenger flow and applied it to study differences between the cities of Chengdu and Qingdao in China. \cite{holleczek2014detecting} used data mining approaches to compare the use of public and private transportation and identify the existence of weak transportation connections.
 
\paragraph{Congestion.} Congestion can dramatically alter travel time estimates for routes that use popular network links in a urban system, but adding network layers in a multilayer transport system can help reduce global congestion \citep{chodrow2016congestion}. Given a network with edges $e$ and flows $j_e$, one can quantify the total time lost in congestion as 

\begin{equation}
    T_c(\textbf{j})=\sum_{e\in \mathcal{L}}j_e(t_e^* - t_e(j_e)),
\end{equation}

where $\textbf{j}$ is the vector of flows whose $e$th element is the flow along edge $e$, $t_e^*$ is the free flow time on edge $e$ (in the absence of congestion), and $t_e(j_e)$ is the congested travel time in the presence of flow $j_e$. 
This measure can be used to analyze the impact of changes in flow along a route. The quantity
\begin{equation}
    \Delta_p = - \nabla T_c(\textbf{j}) \cdot \textbf{e}_p
\end{equation}

where $p$ is a path and $\textbf{e}_p$ is the vector whose $e$th component is 1 iff $e \in p$, quantifies the impact of removing a single unit of flow from $p$ on the global congestion function $T_c$. 
\cite{chodrow2016congestion} quantified how the creation of a planned metro network in Riyadh would affect congestion, by quantifying the change $\Delta_p$ as a function of the speed ratio between the street and metro systems:

\begin{equation}
    \beta = v_{c}/v_{m}
\end{equation}

The authors showed that, as the subway speed increases the global congestion is reduced, but increases locally close to key metro station (Fig.~\ref{Panel2}d and e).

\paragraph{Navigability} 
As cities and their transportation systems become increasingly complex and multimodal, it is important to quantify our difficulty navigating in them. It has been shown that multilayer transport system are characterized by limited navigability, implying that finding one's way is cognitively challenging \citep{gallotti2016limits}. To quantify the difficulty of navigating between two nodes $s$ and $t$ in a network, one can compute the total information value of knowing any of the shortest paths to reach $t$ from $s$:

\begin{equation}
S(s \rightarrow t) = - \log_2 \sum_{\{p(s,t)\}} P[p(s,t)]
\end{equation}

where  ${p(s, t)}$ is the set of shortest paths between $s$ and $t$ (note that there can be more than one with the same length) and $P[p(s,t)]$ is the probability to follow path $p(s,t)$, making the right choice at each intersection along the path \citep{rosvall2005networks}:

\begin{equation}
     P[p(s,t)]=\frac{1}{k_s}\prod_{j\in p(s,t)}\frac{1}{k_j-1}, 
\end{equation}

\cite{gallotti2016limits} quantified the amount of information an individual needs to travel along the shortest path between any given pair of metro stations, in the single-layer metro networks for 15 large cities. They found that this information has an upper bound of the order of 8 bits, corresponding to approximately 250 connections between different routes. Further, studying several among the largest multilayer transport networks (metro/buses/light rail), \cite{gallotti2016limits} showed that the amount of information necessary to know to travel between any two points exceeds the identified cognitive limit of 8 bits in 80\% the cases, suggesting multilayer networks are too complex for individuals to navigate easily (Fig.~\ref{Panel2}c).

\subsubsection{Individual multimodal behaviour}
In recent years, data collected via smart travel cards has dramatically improved our ability to characterise multimodal behaviour in urban transport systems, overcoming some of the limitations related to collecting and analyzing survey data \citep{chen2016promises,zannat2019emerging}. Smart-card automated fare collection systems allow passengers to make journeys involving different transport modes using magnetic cards and automatic gate machines. As these systems identify and store the location and time where individuals board and, in some cases, exit public transport, they collect accurate descriptions of individual travel \citep{pelletier2011smart}. Concurrently, advancements were made possible by the development of methodologies allowing to identify typical travel patterns \citep{ma2013mining}.

\begin{figure}[h!]
    \centering
    \includegraphics[width=\textwidth]{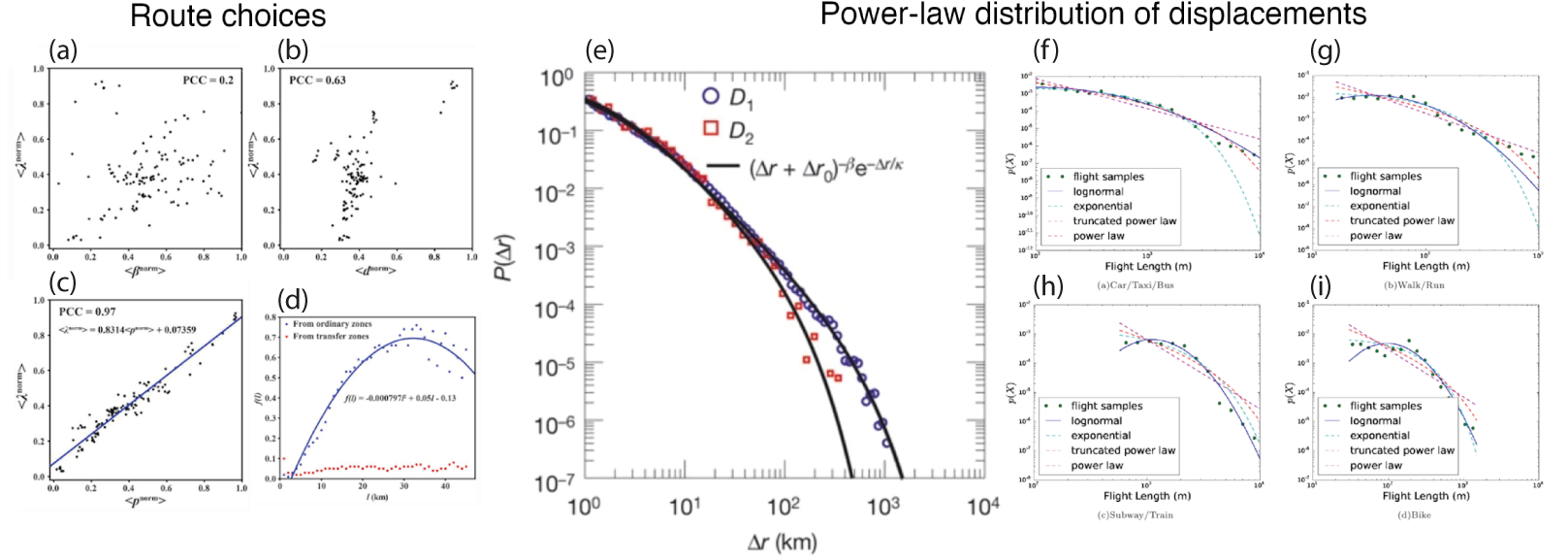}
    \caption{\textbf{Characterizing individual multimodal behaviour.} \textbf{Route choices.} The average coupling $\langle \lambda^{norm} \rangle$ between the bus and metro layers of the public transport system in Shenzen engendered by passengers trips. The quantity is plotted against the speed ratio between the layers $\langle \beta^{norm}\rangle$ (a), the average trip distance $\langle d^{norm} \rangle$ (b), and the fraction of trips originating from nodes serviced by only one mode $\langle p^{norm}\rangle$ (c). All quantities are normalized. The Pearson correlation $PCC$ (reported in each subplots) is significantly positive in subplots (b) and (c), but not in (a), revealing that the coupling between layers is generated by the long-distance trips originating from single-mode nodes, but does not increase with the speed ratio between the layers.  In (d), we show the fraction of multimodal trips against the trip distance $l$, for trips originating from single-mode nodes ('ordinary zones', blue dots) and transfer nodes ('transfer zones', red dots). \textbf{Power-law distribution of displacements.} (e) Probability density function $P(\Delta r)$ of travel distances obtained for two large-scale mobile phone datasets $D1$ and $D2$. The solid line indicates a truncated power law fit $P(\Delta r) \sim \Delta r ^{-\beta}$. \cite{zhao2015explaining} showed that the distribution of travel distances is best described by lognormal distributions for single-mode trips by car (f), foot (g), metro (h) and bicycle (i), and suggest that power-law distributions of travel distances may result from the aggregation of different transport modalities. Figures adapted from \cite{zheng2018coupling,gonzalez2008understanding,zhao2015explaining}.}
    \label{Panel3}
\end{figure}

\paragraph{Route choices.} Smart-card data has allowed to quantify how individuals navigate multilayer networks. One of the key findings is that individuals do not choose optimal paths (those with shortest travel time), especially when the system is congested. Focusing on the bus and subway trips of 2.4 million passengers in Shenzen (China), \cite{zheng2018coupling} studied the coupling (see eq.\ref{eq:coupling}) between the bus and subway layers. In contrast to previous studies \citep{strano2015features}, \cite{zheng2018coupling} characterized the coupling $\lambda$ using passenger behaviour rather than structural properties of the multilayer network. Under their definition, the \emph{coupling} between layers is the fraction of multimodal trips actually undertaken by passengers, rather than the fraction of multimodal shortest paths (see eq.\ref{eq:coupling}). The authors find that this `behavioural' coupling correlates weakly with the empirical speed ratio measured between the two layers over time (Fig.~\ref{Panel3}-a), implying that passengers choose unimodal trips even when multimodal trips may be preferred because one of the two layers is congested. This finding highlights that the speed ratio of different network layers, which was regarded as a key factor in determining coupling strength \citep{strano2015features,chodrow2016congestion}, may have a negligible effect on travelers’ route selections, possibly because passengers do not have a full view of the status of traffic. Instead, \cite{zheng2018coupling} showed that the coupling between layers is generated by long-distance trips originating from nodes served by a single transport layer (Fig.~\ref{Panel3}-b,c and d). 

\paragraph{Power-law distribution of displacements} The availability of large-scale data sources has revealed that individual mobility patterns display universal properties. One key finding is that the distribution $P(\Delta r)$ describing the probability of travelling a given distance $\Delta r$ is characterized by a power-law tail $P(\Delta r)\sim \Delta r ^{-\beta}$, with $1 \leq \beta \leq 2$ \citep{gonzalez2008understanding,barbosa2018human} (Fig.~\ref{Panel3}-e). This finding is consistent across a range of studies that used different data sources, see \cite{alessandretti2017multi} for an extensive review. It was recently shown that these observed scaling properties result from the aggregation of movements within and across characteristic spatial scales, corresponding to the sizes of buildings, neighbourhoods, cities and regions \citep{alessandretti2020scales}. Further, the emergence of scaling properties was associated to the use of multimodal transportation: \cite{zhao2015explaining} used GPS data to show that mobility using a single mode can be approximated by a lognormal distribution, but the mixture of the distributions associated with each modality generates a power-law (Fig.~\ref{Panel3}-f to i); \cite{gallotti2016stochastic} found that a simple model where individual trajectories are subject to changes in velocity generates a distribution of displacements with a power-law tail. In fact, individuals using multimodal infrastructure are subject to drastic changes in velocity. \cite{varga2016further} showed that the travel speed $v$ increases with travel distance according to the power-law functional form $v \sim r ^\alpha$, where $\alpha \approx 0.5$ . This dependence is due to the hierarchical structure of transportation systems and the fact that waiting-times (parking, take-off, landing, etc) decrease as a function of trip distance. 

\section{Open data \& Tools \label{sec:datatools}}

Together with other fields, urban and transportation science are becoming more open, increasingly relying on datasets and computational tools freely available to the scientific community. In this section we highlight some of these openly available datasets and computational resource developed for the analysis of multimodal transport networks.

\subsection*{Data}
During the last years new datasets have been made publicly available either from the public sector or from crowd-sourced data, allowing to go beyond simulations and synthetic data to understand how urban dynamics and mobility unfold, providing a better picture of the world. 

In the last years, the study of transportation networks has benefited from the development of OpenStreetMap \citep{OpenStreetMap}, an open-source collaborative project focused on collecting and sharing world-wide high-quality spatial data \citep{haklay2010openstreetmap,girres2010quality,ferster2019openstreetmap,barbosa2018human}. Data extracted from OpenStreetMap allows to build and analyse several transportation networks, of which the most common are street networks~\citep{boeing2020multiscale,boeing2020world}. Data from OpenStreetMap is further useful to obtain information about infrastructures, such as subways and railways. All such data can be combined to build multimodal transport networks. \cite{gil2015multimodal} obtained multimodal data from OpenStreetMap and built a multiplex network for the Randstad region of the Netherlands linking the layers through the intersections between transport modes. More recently, \cite{natera2019data} followed the same approach to analyse the multiplex transport of fifteen cities in different development stages, including London, Los Angeles, and Mexico City.

As the transportation network of a city also encodes a temporal dimension, it is important to take into account the frequency of buses, tramways, and subways when modeling transportation and mobility. To capture these dynamics, \cite{gallotti2015temporal} constructed and shared the temporal network of public transport in Great Britain. This is a large dataset, where links, associated to flows from one location to another, only exist at specific times. Link information has to be properly combined to compute travel time from origin and destination, highlighting the importance of synchronization of the different transportation models. An interesting feature of this dataset is that it not only contains the public transport layers of several cities, but also connections among them, for instance through coaches, planes, and ferries operating at the national level.

In general, the use of timetables and transit feeds has been enabling researchers to capture with increasing accuracy the dynamics of public transportation systems. A large collection of GTFS feeds in multiple locations has been collected and made freely available as a webpage~\citep{transitfeeds}. These datasets include stops, routes and timetables of public transport in multiple cities and providers from $667$ locations around the world. As shown by \cite{aleta2017transportation}, these data can be used to analyze the public transportation as a multiplex network, considering each bus line and/or transport provider as one layer. Using the data from GTFS feeds,  \cite{kujala2018collection} built and published a collection of 25 urban public transport networks covering cities from North America, Europe, and Oceania. This dataset is peculiar as it also includes the pedestrian layer of the cities, and public transport is further differentiated across the different public transport modes. 

More recently, \cite{tenkanen2020travel} published a very accurate multimodal dataset for the Helsinki region in Finland. This dataset includes multiple transport modes, such as walking, cycling and public transportation options. To calculate the travel times \cite{tenkanen2020travel} use a door-to-door principle. This means that travel time and distance are calculated considering every step of a journey, including walking legs and transfers between vehicles. An important feature of this dataset is the inclusion of travel time matrices for three distinct years, 2013, 2015 and 2018. This is a rare occasion to compare how travel times changed over the years, allowing a characterization of the evolution of human mobility.

Concerning mobility, the Geolife dataset \citep{zheng2011geolife} consists of GPS trajectories collected by Microsoft Research Asia for 178 users in a period of over four years (from April 2007 to October 2011). 69 users labeled their trajectories with the corresponding transportation mode, such as driving, taking a bus, riding a bicycle and walking. As such, the GeoLife data has allowed to investigate mobility behaviours using different transport modes \citep{zhao2015explaining}.

The data described above are freely available, and represent an opportunity for further data-driven investigations of multimodal transportation networks.

\subsection*{Tools}

Over the years computational tools have become more and more important for studying urban systems, and in particular transportation networks. For an overview of the available tools in geographic analysis in transport planning see \citep{lovelace2021open}. 

Multiple tools allow to work with graphs. A few example of freely available softwares and tools are: \textit{Networkx} by \cite{hagberg2008networkx}, \textit{igraph} by \cite{csardi2006igraph}, and \textit{graph-tool} by \cite{peixoto2014graph-tool}. These tools are freely available and constantly updated over time relying on contributions from an engaged community. Although these tools serve a general purpose, they can also be used for the study of transportation networks.

Multiple tools were developed to obtain data on transportation and multimodal infrastructures. One of the best known is \textit{OSMnx} \citep{boeing2017osmnx}, a Python package that downloads street networks from OpenStreetMap into Python objects. \textit{OSMnx} can further be used to download other transportation networks, and build its multimodal transport networks. 

Another reliable Python library to read data from OpenStreetMap and extract transportation networks is \textit{Pyrosm} \citep{tenkanen2020pyrosm}. Differently from \textit{OSMnx}, \textit{Pyrosm} reads the data directly from OpenStreetMap's Protocol Buffer Format files (*.osm.pbf), while OSMnx downloads the data from the Overpass API~\citep{overpass}. For this reason \textit{Pyrosm} is a particularly good alternative when working with large urban areas, states, and even countries, while OSMnx typically offers a more precise way to collect data from specific points in a city.

To work with public transportation data \cite{google2020gtfs} developed \textit{transitfeed}, a Python library to parse, validate and build GTFS files. This tool is particularly useful to those interested in the manipulation of the raw data. However, to convert the data into a network, some additional steps are needed. An alternative to read the GTFS feeds and directly extract its transportation network is \textit{Peartree} \citep{butts2021peartree}, a Python library allowing to convert GTFS feed schedules into the corresponding directed network graph.

\textit{Movingpandas}, developed by \citep{graser2019movingpandas} is a Python package that provides trajectory data structures and functions for the analysis and visualisation of mobility data. In a similar sense, and also developed in Python, \textit{scikit-mobility}~\citep{pappalardo2021scikitmobility} is a library that implements a framework for analyzing statistical patterns and modeling mobility, including functions for estimating movement between zones using spatial interaction models, and tools to asses privacy risks related to the analysis of mobility datasets.

The aforementioned tools were not built specifically with the purpose to work with multiplex networks. To cover this need, specific libraries have been developed. A first example is \textit{muxViz} by \cite{dedomenico2015muxviz}, a stand-alone front-end tool which allows the computation of several multilayer measures, from centrality to community detection. \textit{MuxViz} is also an advanced visualization tool, providing an effective way to display edge-colored multigraph or multislice networks. 

Several software options are available in Python, often built on top of NetworkX. A library originally designed for the study of multilayer networks, that can be easily adapted to multimodal networks, is \textit{MAMMULT} by \cite{nicosia2015mammult}. This library contains a collection of algorithms to analyze and model multilayer networks. The functions included in the collection cover a wide range of applications from structural properties, such as node, edge, and layer basic properties, to the analysis of dynamics on multilayer networks, such as random walks. 

Another example is \textit{multiNetX} by \cite{kouvaris2015pattern}. This library extends Networkx allowing the creation of undirected weighted and unweighted multilayer networks from Networkx objects. Once the multilayer networks are built, the library focuses on the spectral properties of the corresponding adjacency or Laplacian matrices. Such tool also provides nice visualization tools improving from Networkx, allowing the user to better visualize multilayer dynamics through coloring the nodes and links over time. 

A more recently developed Python library, not relying on Networkx, is \textit{Pymnet} \citep{kivela2018pymnet}. The package handles general multilayer networks, including multiplex networks with temporal variables. For this reason, it is possible to use it for the analysis of multimodal urban transport networks that incorporate transit schedules. This library also includes multiple network analysis methods, transformations, and models to analyze and visualize multilayer networks. Another alternative is the \textit{multinet} library \citep{magnani2020multiplex}, available both in Python and R. This package provides tools to work with multilayer networks, including community detection and visualizations. When visually working with multilayer networks it is important to account for principles of visualization and cognitive overload \citep{rossi2015towards}. Another option in R is \textit{multiplex} developed by \cite{rivero2020algebraic}. This library offers multiple functions to work with matricial representations and visualization of multilayer networks. 

Finally, tools such as \textit{mapbox}~\citep{mapbox}, \textit{carto}~\citep{carto}, \textit{kepler.gl}~\citep{kepler2021kepler}, and \textit{studio unfolded}~\citep{unfolded}, built on top of OpenStreetMap~\citep{OpenStreetMap}, allow to create and share geospatial interactive web visualizations. While these tools have not been designed to work specifically with networks, it is possible to leverage their geospatial visualization capabilities to create appealing visualization of urban systems.

All the aforementioned tools are available freely online, with an open source code open to edits, collaborations and improvements. Most tools for multilayer network analysis currently serve a general purpose and have not been designed to support features for multimodal transportation networks in particular. Given the growing interest in this topic, we anticipate future open source libraries built specifically for multimodal urban data.

\section{Conclusions \label{sec:conclusions}}

In this review we discussed the state-of-the art in the field of multimodal mobility and multilayer transport networks from a complexity science perspective, focusing on urban environments. On one hand we covered the science of the \emph{dynamics} of mobility: How do people move? Which forms of transportation do they use? How do they find their paths or switch between modes? On the other hand these dynamics take place on an underlying (infra)\emph{structure} which can be well modeled by multilayer networks. In this context, a number of mathematical metrics have been developed in network science recently which allow the rigorous study of the topic. Parallel to the methodological developments we have witnessed a spur of new computational tools -- many of them open source -- and datasets which considerably facilitate and boost further research on the topic. Despite an explosion in geospatial data collection, it is still relatively difficult to access spatio-temporally fine-grained -- and appropriately anonymized \citep{de2013unique} -- mobility data openly. Such high-quality data are in danger of being siloed in by commercial stakeholders, obstructing transparent research on the topic. We must therefore push for the implementation of better systems by governments, academia, and industry to recognize and promote efforts for making data sets and tools openly available by and for researchers \citep{stodden2016enhancing,lovelace2021open}. 

The increasing availability of mobility data, concurrently with the continuous growth and developments of urban transport infrastructures are raising new research challenges. A first critical issue relates to the modelling of shared mobility services, such as shared bicycles and vehicles \citep{shaheen2016mobility}. Multimodal frameworks that integrate shared services with traditional public and private transport infrastructures are becoming necessary to ensure real-time and user-centered solutions for planning, forecasting and managing services, while increasing safety, reducing congestion and emissions. A second important area focuses on investigating at scale the decision-making processes underlying individual transport mode choice and routing behaviour leveraging the increasingly available high-resolution individual traces. We anticipate that this new understanding will be key to describe how microscopic decision-making processes contribute to the emergence of collective mobility flows in multimodal systems.  

Despite the currently exploding research on multimodal mobility, there exists a wide frontier of topics to tackle and new approaches to explore. Important advances on multimodal mobility and transportation have been shown to be interdisciplinary, and have clearly benefitted from the large variety of scientific fields and practices. Indeed, synergies between disciplines such as urban planning, geoinformatics, computer science, physics, etc., have increased in the last few years, giving rise to new interdisciplinary approaches such as a Science of Cities or Urban Data Science~\citep{boeing2021urban,resch2019hds}. We envision that research on multimodal mobility and transportation to maintain a highly interdisciplinary character also in the future. For example, the study of human mobility has recently benefited from novel scientific advances from other fields such as deep learning. \cite{luca2020deep} offers a comprehensive overview of the topic and its applications to human mobility, surveying data sources, public datasets, and deep learning models, and we anticipate the possibility that this area will soon make an impact in unveiling new features of multimodal mobility and transportation. 

Finally, understanding multimodal mobility and its underlying infrastructure is of paramount importance for developing sustainable urban transport, as it relies on the central role of public transport modes. Indeed, studying multimodal mobility is one piece in the puzzle towards reversing the global societal threat of climate change which is caused to a considerable extent by car-centric transport monocultures \citep{mattioli2020political}. For this reason, we hope that our review can serve as a starting point to develop a more modern, sustainable and integrated idea of mobility worldwide.

\small{
\setlength{\bibsep}{0.00cm plus 0.05cm} % no space between items
\bibliographystyle{apalike}
\bibliography{bib_multimodality.bib}
}

\end{document}